\documentclass[16pt, usenatbib]{mn2e}
\usepackage{graphicx,times, subfigure}
\usepackage{bm}
\usepackage{epsf}
\usepackage{amsmath}
\usepackage{amssymb}
\usepackage{cases}
\usepackage{color}

\def\cb{\textcolor{black}}
\def\la{\langle}
\def\ra{\rangle}

\def\be{\begin{equation}}
\def\ee{\end{equation}}
\def\ben{\begin{eqnarray}}
\def\een{\end{eqnarray}}
\def\nn{\nonumber}
\def\oh{\hat\Omega}
\def\myC{{\cal C}}

\def\myf{\Psi}
\def\rB{\rm B}
\def\rR{\rm R}
\begin{document}
\onecolumn
\title[Probing Modified Gravity Theories with ISW and CMB Lensing]
{Probing Modified Gravity Theories with ISW and CMB Lensing}
\author[Munshi et al.]
{D. Munshi$^{1}$, B. Hu$^{2}$, A. Renzi$^{3,4}$, A. Heavens$^{5}$, P. Coles$^{1}$\\
$^{1}$ Astronomy Centre, School of Mathematical and Physical Sciences, University of Sussex, Brighton BN1 9QH, U.K.\\
$^{2}$ Instituut-Lorentz Theoretical Physics, Universiteit Leiden, Niels Bohrweg 2, 2333 CA
Leiden, The Netherlands\\
$^{3}$ Department of Mathematics, University of Rome Tor Vergata\\
$^{4}$INFN, Sezione di Roma Tor Vergata, Rome, Italy\\
$^{5}$ Imperial Centre for Inference and Cosmology, Department of Physics,  Imperial College, Blackett Laboratory, Prince Consort Road, London SW7 2AZ, U.K., \\}
\maketitle
\begin{abstract}
We use the optimised skew-spectrum as well as the skew-spectra associated with the Minkowski Functionals (MFs) to 
test the possibility of using the  cross-correlation of the Integrated Sachs-Wolfe effect (ISW) and 
lensing of the cosmic microwave background (CMB) radiation
to detect deviations in the theory of gravity away from General Relativity (GR).
We find that the although both statistics can put constraints on modified gravity, the optimised skew-spectra are especially sensitive to the parameter $\rB_0$ 
that denotes the the {\em Compton wavelength} of the scalaron at the present epoch.
We investigate three modified gravity theories, namely: the Post-Parametrised Friedmanian (PPF) formalism;
the Hu-Sawicki (HS) model; and the Bertschinger-Zukin (BZ) formalism.
Employing a likelihood analysis for an experimental setup similar to ESA's Planck mission, we find that, assuming GR to be the correct model, 
we expect the constraints from the first two skew-spectra, $S_{\ell}^{(0)}$ and 
$S_{\ell}^{(1)}$, to be the same: $\rm B_0<0.45$ at $95\%$ confidence level (CL),  
and $\rm B_0<0.67$ at $99\%$ CL in the BZ model. The third skew-spectrum does not
give any meaningful constraint. 
We find that the optimal 
skew-spectrum provides much more powerful constraint, giving $\rm B_0<0.071$ at $95\%$ CL and $\rm B_0<0.15$ at $99\%$ CL,
which is 
essentially identical to what can be achieved using the full bispectrum.
\end{abstract}
\begin{keywords}: Cosmology, Methods: analytical, statistical, numerical, modified gravity, dark energy
\end{keywords}

\section{Introduction}
\label{sec:intro}
The observations of type Ia supernovae imply that our Universe is undergoing 
a phase of accelerated expansion \citep{Riess98,Perl99}. 
Cosmic acceleration can arise from either an exotic form of energy with negative pressure, 
referred to as ``dark energy", or a modification of gravity manifesting on large scales. 
As shown by various authors \citep{Bertschinger:2006aw,Song:2006ej,Brax:2008hh,2013arXiv1312.5742H}, determining the cause of the acceleration os hampered by the fact that the 
background dynamics in dark energy and modified gravity models are 
nearly indistinguishable. To lift this degeneracy, one can test the evolution of perturbations in these models. 
The perturbative approach to growth of structure in modified gravity can, in principle, be classified in two different frameworks: parametric and non-parametric, an example of the latter being 
principal component analysis~\citep{Zhao:2008bn,Zhao:2009fn,Zhao:2010dz,Hojjati:2011xd}. In this paper we focus on the former.

There exist several phenomenological parametrizations of modified gravity including
the Bertschinger-Zukin \citep{BZ08} parametrization, and that of \cite{Staro07}). 
These parametrizations are suitable for the quasi-static regime, 
where the time evolution of the gravitation potentials is negligible compared with their spatial gradient. 
Furthermore, if we focus on the linear fluctuation dynamics for which the equations in Fourier space can be reduced to simple algebraic relations, 
these techniques allow us to perform some analytic calculations which make the parametrization technically efficient. 
However, if we want to go further beyond the quasi-static scale, while remaining in the linear perturbation framework, 
the parametrization of modified gravity becomes more complex. 
This is because on the largest scales, especially the super- or near-horizon scales, 
the time evolution of the gravitational potentials is no longer negligible.  In fact, the time derivative terms dominate the dynamical equations, 
which means that we need to solve some temporal ordinary differential equations. 
All in all, the inclusion of time derivative terms makes the parametrization of modified gravity not so manifest anymore 
Actually, there exists some debate about the range of validity of the 
various parametrizations; on the one hand, as shown by \cite{Zuntz:2011aq}, 
using a parametrization with insufficient freedom significantly tightens the apparent theoretical constraints. 
On the other hand, for some specific modified gravity models some phenomenological parametrizations work quite well; 
for instance \cite{Hojjati:2012rf} recently demonstrated that for 
small Compton wavelength in the $f(\rR)$ model, 
the Bertschinger-Zukin parametrization is in practice good enough for current data analysis.
This is because, for small Compton wavelengths, the most significant modifications w.r.t. 
GR occur in the sub-horizon regime, while the modification on the super-horizon scales are subdominant. 
In addition to the above explicit parametrizations, some quite generic  frameworks have been proposed, 
such as the parametrized Post-Friedmann (PPF) formalism, including the Hu-Sawicki approach \citep{Hu:2007pj,Fang:2008sn}, its calibration version 
\citep{Lombriser:2013aj} and Baker-Ferreira-Skordis-Zuntz algorithm \citep{Baker:2011jy,Baker:2012zs}, 
and the Effective Field Theory (EFT) formalism \citep{Gubitosi:2012hu,Bloomfield:2012ff,2013arXiv1312.5742H}. 
These formalisms are devoted to build up a ``dictionary'' of modified gravity theories and their PPF or EFT correspondence.
Since the purpose of these generic formalisms is to construct a unified way to include all the modified gravity/dark energy models, 
they contain more arbitrary functions/coefficients, which usually lead to looser constraints.

Besides the recent progress on the construction of parametrizations, many observational windows have recently been proposed, 
such as the Integrated Sachs-Wolfe (ISW) effect~\citep{Sachs:1967er} in Cosmic Microwave Background (CMB) anisotropies~
\citep{Zhang:2005vt,Song:2007da,2008PhRvD..78d3519H}, the power spectrum of 
luminous red galaxies~\citep{2010PhRvD..81j3517Y,2012PhRvD..86j3505H,2013PhRvD..88d4050A}, 
cluster abundance~\citep{Jain:2007yk,2009PhRvD..80h3505S,Lombriser:2010mp,2011PhRvD..83f3503F}, 
Coma cluster \citep{2013arXiv1312.5083T},
galaxy peculiar velocities~\citep{Li:2012by}, redshift-space distortions~\citep{Jennings:2012pt,2013MNRAS.436...89R}, 
weak-lensing~\citep{Heavens2007,Zhang:2007nk,2010Natur.464..256R,2008PhRvD..78d3520H,2010PhRvD..81l3508D,2011A&A...530A..68T,2012MNRAS.423.1750L,2013MNRAS.429.2249S}, 
$21$cm observations~\citep{Hall:2012wd}, matter bispectrum~\citep{GilMarin:2011xq,2013JCAP...03..034B}, {\it etc}. 
In addition, recently some N-body simulation algorithms in modified gravity models have been developed~\citep{Zhao:2010qy,Li:2010zw}. 
As shown by \cite{Song:2007da} and \cite{Lombriser:2010mp}, with WMAP resolution the modification effects on the CMB mainly come from the ISW effect, 
which becomes prominent on the super-horizon scales. However, due to the unavoidable cosmic variance on large scales, 
the constraints from these effects are not significant. 
On the other hand, since the typical modification scales are on sub-horizon scales, 
several studies show that the most stringent constraints come from the large-scale structure data sets. 
For example, the strongest current constraint on $f(\rR)$ gravity (${\rm log}_{10}{\rm B}_0<-4.07;\ 95\%{\rm CL}$)~\citep{2014arXiv1401.3980D} 
is driven by the galaxy spectrum from WiggleZ data sets~\citep{2012PhRvD..86j3518P}.
Various previous results show that the main constraint on modified gravity comes from galaxy or cluster scales which corresponds 
to the multipole range $\ell\gtrsim 500$ in CMB data, where lensing 
effects are no longer negligible. The recent release of {\it Planck} 
data \citep{Ade:2013ktc} provides us with a fruitful late-time information both on ISW and lensing, 
which is encoded in the CMB temperature power-spectrum 
\citep{Ade:2013kta}, the lensing potential power-spectrum \citep{Ade:2013tyw}, and the CMB temperature ISW-lensing bispectrum \citep{Ade:2013ydc,Ade:2013dsi}. 
The full sky lensing potential map has been constructed and the amplitude of the lensing potential power-spectrum 
has been estimated at the $25\sigma$ level. 
The ISW-lensing bispectrum is also detected with nearly $3\sigma$ confidence level. 
Although the ISW-lensing bispectrum data have not yet been released, forecasts of constraints on modified gravity models through this novel observational statistic have been 
investigated \citep{DiValentino:2012yg,HLBM12}. 
These studies show that the ISW-lensing bispectrum is an effective tool to constrain modified gravity. 
Also notice that  \cite{Hu13} analysed CMB temperature power-spectrum data
alone and improved the previous constraint from WMAP9's ${\rm B}_0<3.37$ at $95\%$ CL to ${\rm B}_0<0.91$. 
Inclusion of the lensing potential power spectrum improved it to ${\rm B}_0<0.12$. The lensing-ISW bispectrum is known to be uncorrelated to the power-spectrum and thus
it can further tighten the constraint on ${\rm B}_0$.

Inspired by these results, in this paper we use the recently introduced optimum skew-spectra and the skew-spectra associated 
with the Minkowski Functionals (MFs) to constrain departures from GR.
Since their introduction in cosmology by \cite{MBW94}, MFs have
been extensively developed as a statistical tool for non-Gaussianity
in a cosmological setting for both two-dimensional (projected) and
three-dimensional (redshift) surveys. Analytic results are known certain
properties of the MFs of  a Gaussian random field making them
suitable for identifying non-Gaussianity. Examples of such studies
include CMB data \citep{Schmalzing98,Novikov00,HikageM08,Natoli10},
weak lensing (\cite{Matsubara01,Sato01,Taruya02,MuWaSmCo12}),
large-scale structure
\citep{Gott86,Coles88,Gott89,Melott89,Gott90,Moore92,Gott92,Canavezes98,SSS98,
Schmalzing00,Kerscher01,Hikage02,Park05,HKM06,Hikage08}, 21cm
\citep{Gleser06}, frequency cleaned Sunyaev-Zel'dovich (SZ) maps
\citep{MuSmJoCo13} and N-body simulations
\citep{Schmalzing00,Kerscher01}. The MFs are spatially-defined
topological statistics and, by definition, contain statistical
information of all orders in the moments. This makes them
complementary to the poly-spectra methods that are defined in
Fourier space. It is also possible that the two approaches will be
sensitive to different aspects of non-Gaussianity and systematic
effects, although in the weakly non-Gaussian limit it has been shown
that the MFs  reduce to a weighted probe of the bispectrum
\citep{HKM06}.

The skew-spectrum is a weighted statistic that can be tuned to a
particular form of non-Gaussianity, such as that which may arise
either during inflation at an early stage or from structure
formation at a later time. The skew-spectrum retains more
information about the specific form of non-Gaussianity than the
(one-point) skewness parameter alone. This allows not only the
exploration of primary and secondary non-Gaussianity but also the
residuals from galactic foreground and unresolved point sources. The
skew-spectrum is directly related to the lowest-order cumulant
correlator and is also known as the two-to-one spectra in the
literature \citep{Cooray01}. In a series of recent publications the
concept of skew-spectra was generalized to analyse the morphological
properties of cosmological data sets or in particular the MFs
\citep{MuSmCooReHeCo13, MuWaSmCo12,MuSmJoCo13,PratMun12}. The first of these
three spectra, in the context of secondary-lensing correlation
studies, was introduced by \cite{MuVaCoHe11} and was subsequently
used to analyse data release from WMAP  by \cite{Cala10}.

The layout of the paper is as follows. In \textsection\ref{sec:mod_grav}
we briefly outline various models and parametrization of modified gravity.
Next, in \textsection\ref{sec:isw_lensing}, we review the
non-Gaussianity, at the level of bispectrum, introduced by cross-correlaion of secondaries and lensing of CMB. 
In \textsection\ref{sec:MF} we introduce the skew-spectra associated with the Minkowski Functionals (MFs) and compute them for
various modified gravity scenarios. \textsection\ref{sec:like} is devoted to likelihood analysis using MFs.
In \textsection\ref{sec:result} we discuss our results. 
Finally \textsection\ref{sec:disc} is reserved for concluding remarks as well as discussing the future prospects. 

\section{Modified gravity Models}
\label{sec:mod_grav}
Studies of modified gravity models can, in principle, be classified into two different frameworks \citep{BZ08}. 
The first is a model-dependent method. One can start from a specific Lagrangian, investigating its dynamical behaviour to finally give its predictions. Various viable modified gravity models have been proposed  which fall into this category \citep{Clifton:2011jh}. In this paper we mainly focus on $f(\rR)$ models (see e.g. \cite{DeFelice:2010aj} for a review), such as the \cite{Starobinsky:1980te} model.
or the Hu-Sawicki model \citep{HuSaw07}. 

The other method is inspired by the parametrized Post-Newtonian (PPN) approach to solar-system tests of gravity. 
In this case one aims to build a
model-independent framework, in which many modified gravity models can be parametrized in a unified way.
The simplest idea is directly to generalize the Eddington parameter ($\gamma\equiv\Phi/\Psi$; \cite{Edd}) to an unknown function of space and time $\gamma(t,{\bf x})$ in a Friedmann Universe.
Many studies, such as \citep{BZ08,Zhao:2008bn,
Zhao:2009fn,Hojjati:2011ix,Giannantonio:2009gi} show that this works quite well for large-scale structure data. This is because these parametrizations are mainly suitable for the quasi-static regime where the time evolution of the gravitational potentials are negligible compared with spatial gradients. Furthermore, if we focus on the linear 
analysis in the Fourier domain, then the dynamical equations can be reduced to simple algebraic relations. 
These allow us to perform some analytic calculations, which make the parametrization technically efficient. 
However, if we want to go further, beyond the quasi-static scale, even though  still in the linear regime, the parametrization of modified gravity is more non-trivial. This is because at the larger scales, especially the super- or near-horizon scales, the time evolution of gravitational potentials is no longer negligible and we need to solve temporal ordinary differential equations. 


Beside the above explicit parametrizations, some quite generic frameworks have been proposed, such as the Hu-Sawicki parametrized Post-Friedmann (PPF) formalism \citep{Hu:2007pj,Hu:2008zd,Fang:2008sn} and its calibration version \citep{Lombriser:2013aj}. The Hu-Sawicki PPF parametrization is defined by three functions: $g(\ln a,k_H), f_{\zeta}(\ln a), f_G(\ln a)$ and a single parameter $c_{\Gamma}$. They correspond to the metric ratio, the super-horizon relationship between the metric and density, the deviation of Newton's constant on super-horizon scale from that on quasi-static scales, and the relationship between the transition scale and the Hubble scale~\citep{Hu:2007pj}. Of course, this formalism is quite generic. However, in order to obtain the explicit parametrization form of these arbitrary functions, one needs to solve the exact equation of motion obtained from the original Lagrangian of the modified gravity theory and fit the above three functions with the exact solution.
Up to now, only a few models, such as $f(\rR)$ and DGP models, have been successfully implemented in the the Hu-Sawicki PPF formalism. Even though, this formalism still has a great advantage for numerical purposes, since it provides a unified form to write down all the modified equations.
Besides what mentioned above, there exist many other parametrizations \citep{Bean:2010zq,Bertacca:2011wu,Linder:2005in,
Gubitosi:2012hu,Bloomfield:2012ff,Baker:2011jy,Baker:2012zs,Amendola:2007rr,Brax:2012gr}. 

\subsection{Hu-Sawicki $f(\rR)$ model}
As an example of a model-dependent method, the Lagrangian of Hu-Sawicki model (hereafter HS) reads:
\ben\label{Hu-Sawicki}
f(\rR)=-m^2\frac{c_1(R/m^2)^n}{c_2(R/m^2)^n+1}\;; \quad\quad
m^2\equiv H_0^2\Omega_{m}=(8315{\rm Mpc})^{-2}\left(\frac{\Omega_m h^2}{0.13}\right)\;.\een
As shown by \cite{HuSaw07}, this model can pass the local solar system tests.
The non-linear terms in $f(\rR)$ introduce fourth-order derivatives into this theory, rather than the more familiar second-order derivatives. Fortunately, we can reduce the derivatives to second order by
defining an extra scalar field $\chi\equiv ({\rm d} f/{\rm d} R)$, namely the ``scalaron'', which absorbs the higher derivatives.   
The {\em Compton wavelength} of the scalaron is defined as
\ben\label{comp_wav}
\rB=\frac{f_{\rR\rR}}{1+f_{\rR}}\rR'\frac{H}{H'}\;,\een
with $f_R={\rm d} f/{\rm d} \rR$, $f_{RR}={\rm d}^2 f/{\rm d} \rR^2$ and ${~}'\equiv{\rm d} /{\rm d} \ln a$.
In the high curvature regime, Eq.(\ref{Hu-Sawicki}) can be expanded w.r.t. $(m^2/\rR)$ as:
\ben\label{Hu-SawickiApp}
\lim_{(m^2/R) \rightarrow 0}f(\rR)\approx -\frac{c_1}{c_2}m^2+\frac{c_1}{c_2^2}m^2\left(\frac{m^2}{\rR}\right)^n+\cdots\;.\een
From Eq.(\ref{Hu-SawickiApp}) we can see that, the first and second terms represent a cosmological term and a deviation from it, respectively.  In order to mimic $\Lambda$CDM evolution on the background, the value of $(c_1/c_2)$ can be fixed ~\citep{HuSaw07} such that:
$({c_1}/{c_2})=6(\Omega_{\Lambda}/\Omega_m)$. By using this relation the number of free parameters can be reduced to two. 
From the above analysis, we can see that, strictly speaking, due to the appearances of correction terms to the cosmological constant, the HS model cannot exactly mimic $\Lambda$CDM. Since $(m^2/\rR)$ increases very fast with time, the largest value (at the present epoch) is $(m^2/\rR)\sim0.03$, the largest deviation to the $\Lambda$CDM background happens when $n=1$, with $1\%$ errors, corresponding to $(m^2/\rR)c_2\sim 0.01$ in Eq.(\ref{Hu-SawickiApp}). For larger $n$ values, such as $n=4,\ 6$ we can safely neglect this theoretical error. As shown by \cite{HLBM12}, for $n=1$ this $1\%$ deviation from $\Lambda$CDM  brings a $10\%$ error in the variance of the parameter $\rB_0$, while for $n=4,6$ our results are not affected. 

Without loss of generality, we can choose the two free parameters to be ($n,c_2$). However, for more general $f(\rR)$ models the $\Lambda$CDM evolution of the background can be reproduced exactly by only introducing one free parameter \citep{Song:2006ej}. 
This means that there exists some degeneracy between the two parameters. Usually General Relativity (GR) is recovered when $\rB_0=0$.
As demonstrated by \cite{HLBM12}, no matter what value $n$ takes, we are always allowed to set 
$\rB_0=0$ by adjusting $c_2$. Furthermore, in order to mimic $\Lambda$CDM on the background, $c_2$ and $n$ need to satisfy one constraint: the first term in the denominator of Eq.(\ref{Hu-Sawicki}) should be much larger than the second. This condition gives:
\begin{subnumcases}{\rB_0^{{\rm max}}=}
  0.1\;,\quad (n=1)\;,\nonumber\\
  1.2\;,\quad (n=4)\;,\label{B0Max}\\
  4.0\;,\quad (n=6)\;.\nonumber
\end{subnumcases}

\cite{HLBM12} forecast that Planck\footnote{http://sci.esa.int/science-e/www/area/index.cfm?fareaid=17} is expected to reduce the error bars on the modified gravity parameter $B_0$ by at least one order of magnitude compared to WMAP. The spectrum-bispectrum joint analysis can further improve the results by a factor ranging from $1.14$ to $5.32$ depending on the value of $n$.

\subsection{Hu-Sawicki PPF formalism (PPF)}
In contrast to the above subsection, in what follows we will consider all possibilities in $f(\rR)$ gravity which can mimic 
the $\Lambda$CDM background in the Hu-Sawicki PPF formalism (hereafter PPF).
The logic of the PPF formalism is the following: first, considering two limits in the linear fluctuation regimes, the super-horizon and quasi-static regimes. In the former the time derivatives are much more important than the spatial derivatives and in the latter limit the vice versa; then derive and solve the gravitational equations in these limits. Given the knowledge of these two limits, one can propose two modified gravitational equations which recover the above results in the super-horizon and quasi-static limits, respectively. Finally, we integrate all the linear scales using the proposed equations.

For the metric scalar fluctuations, in principle we have only two degrees of freedom, such as $\Phi$ (Newtonian potential) and 
$\Psi$ (curvature potential)  in the conformal Newtonian gauge, which means we only need two dynamical equations. For PPF, these two master equations are the modified Poisson equation and the equation for $\Gamma$:
\ben
&& k^2\Big[\Phi_-+\Gamma\Big]=4\pi Ga^2\rho_m\Delta_m\;;  \quad\quad \Phi_{-}= \Phi-\Psi\label{eq:Master1}\\
&& (1+c_{\Gamma}^2k_H^2)\Big[\Gamma'+\Gamma+c_{\Gamma}^2k_H^2(\Gamma-f_{G}\Phi_-)\Big]=S. \label{eq:Master2}\een
Where the source term $S$ is given by:
\begin{eqnarray}\label{source}
&& S=-\left[\frac{1}{g+1}\frac{H'}{H}+\frac{3}{2}\frac{H_m^2}{H^2a^3}(1+f_{\zeta})\right]\frac{V_m}{k_H}
+\left[\frac{g'-2g}{g+1}\right]\Phi_-\;.\end{eqnarray}
$V_m$ here is the scalar velocity fluctuation of the matter in both the comoving and Newtonian gauge.
and $H_m$ is the contribution to Hubble parameter from the matter component; see \cite{Hu:2007pj} for more details.

In Eq.(\ref{eq:Master2}), the coefficient $c_{\Gamma}$ represents the relationship between
the transition scale and the Hubble scale, and the function $f_{\zeta}$ gives the relationship between
the metric and the density perturbation. For $f(\rR)$ models, we have  $c_{\Gamma}=1$, $f_{\zeta}=c_{\zeta}g$ and
the function $g(\ln a,k)$ can be expressed as follows:
\begin{equation}\label{g_interplt}
g(\ln a,k)=\frac{g_{{\rm SH}}+g_{{\rm QS}}(c_gk_H)^{n_g}}{1+(c_gk_H)^{n_g}}\;,\quad\quad
g_{{\rm QS}}=-1/3\;,\quad n_g=2\;,\quad c_g=0.71\sqrt{B(t)}\;.
\end{equation}
The above descriptions have been implemented in the publicy-available PPF module \citep{Fang:2008sn} of {CAMB}\footnote{http://camb.info/}~\citep{Lewis:1999bs}. 
The current constraints on general $f(\rR)$ models within the Hu-Sawicki PPF formalism are $\rB_0<0.42 (95\%{\rm CL})$ 
by using CMB and ISW-galaxy correlation data, and a strong constraint $\rB_0<1.1\times10^{-3}$ at $95\%$ CL \citep{Lombriser:2010mp}. 
using a larger set of data, such as WMAP5\footnote{http://map.gsfc.nasa.gov/}, ACBAR\footnote{http://cosmology.berkeley.edu/group/swlh/acbar/}, 
CBI\footnote{http://www.astro.caltech.edu/~tjp/CBI/}, VSA, Union\footnote{http://supernova.lbl.gov/Union/}, 
SHOES, and BAO data. 

\subsection{Bertschinger-Zukin formalism (BZ)}
Another popular phenomenological parametrization was proposed by \citep{BZ08} (hereafter BZ) and implemented in the Einstein-Boltzmann solver {MGCAMB}\footnote{http://www.sfu.ca/~aha25/MGCAMB.html} \citep{Zhao:2008bn,Hojjati:2011ix}.
The logic of this parametrization is to re-write the two gravitational potentials in terms of two observation-related variables, the time- and scale- dependent Newton constant $G \mu(a,k)$ and the so-called gravitational slip $\gamma(a,k)$:
\begin{eqnarray}
\label{BZ}
k^2\Psi = -4\pi Ga^2\mu(a,k)\rho\Delta; \quad\quad
\frac{\Phi}{\Psi}&=&\gamma(a,k).
\end{eqnarray}
$G$ is the Newton constant in the laboratory.
Furthermore, in the quasi-static regime, Bertschinger and Zukin propose a quite efficient parametrizations for these two quantities (see also \cite{Zhao:2008bn}):
\begin{eqnarray}
\label{BZ2.1}
\mu(a,k) = \frac{1+\frac{4}{3}\lambda_1^2k^2a^4}{1+\lambda_1^2k^2a^4}; \quad\quad\quad
\gamma(a,k) = \frac{1+\frac{2}{3}\lambda_1^2k^2a^4}{1+\frac{4}{3}\lambda_1^2k^2a^4}\;.
\end{eqnarray}
The above parametrization was refined to take the ISW effect into account through an empirical formula \citep{Giannantonio:2009gi}:
\ben
\label{BZ3}
&& \mu(a,k) = \frac{1}{1-1.4\times10^{-8}|\lambda_1|^2a^3}\left [ \frac{1+\frac{4}{3}\lambda_1^2k^2a^4}{1+\lambda_1^2k^2a^4}\; \right ].
\een
Compared with PPF, one can easily see the physical meaning of parameter $\lambda_1$, as the present Compton wavelength $\lambda_1^2=\rB_0c^2/(2H_0^2)$. Beside that, we can also see that BZ is much more efficient than the former, because in BZ one only needs to solve an algebraic relation, Eq.(\ref{BZ}) or equivalently Eq.(\ref{BZ3}), while in PPF we have to integrate differential equations, Eq.(\ref{eq:Master1}) and Eq.(\ref{eq:Master2}). The price BZ pays is that it might not account for the ISW effect properly in the super-horizon regime. However, recently it was shown 
\citep{Hojjati:2012rf} that for all practical purposes BZ for $f(\rR)$ model with small $\rB_0$ is good enough even if one considers the near-horizon scale: the maximum error is $\mathcal O(2\%)$. Recently it was shown
by \citep{HLBM12} that the temperature and lensing power spectrum data from Planck alone can give an upper bound on $\rB_0<0.91$ at $95\%$CL

\section{ISW-Lensing cross-spectra as a probe of Modified Gravity Theories}
\label{sec:isw_lensing}
\begin{figure}
\begin{center}
{\epsfxsize=10 cm \epsfysize=8 cm {\epsfbox[31 330 589 716]
{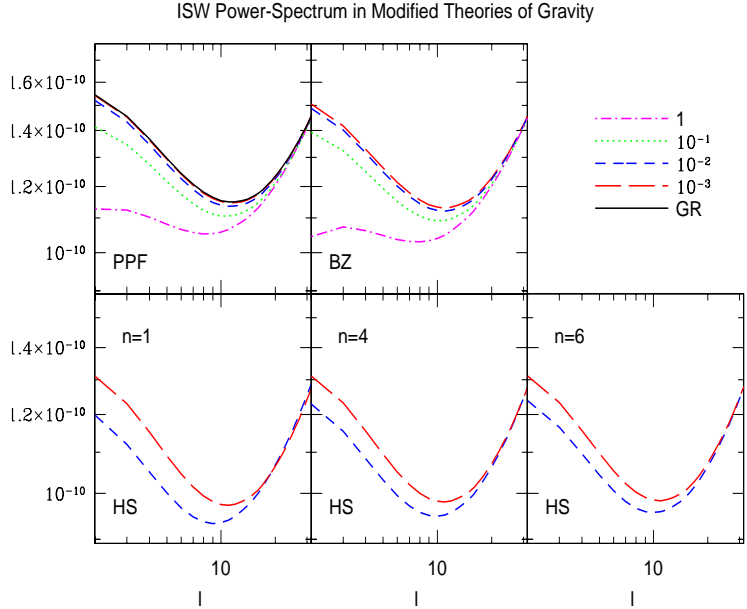}}}
\end{center}
\caption{The ISW contribution to the (dimensionless) temperature power spectrum $\ell(\ell+1)\myC_{\ell}^{\rm TT}/2\pi$ is depicted as a function of the parameter ${\rm B}_0$ of the PPF formalism. The general relativisitic (GR) predictions correspond to ${\rm B}_0=0$ (top-left panel). 
The top-left and top-right panels correspond to PPF (top-left)  and BZ (top-right) formalism.
For the PPF we chose ${\rm B}_0 = 1, 10^{-1}, 10^{-2}, 10^{-3}$. 
The bottom panels correspond to the predictions from HS \citep{HuSaw07} with $n=1$ (bottom-left), $n=4$ (bottom-middle) and $n=6$ (bottom-right).
The values of ${\rm B_0}$ in these plots are ${\rm B}_0 = 10^{-2}$ and ${\rm B}_0 =10^{-3}$.}
\label{fig:clTT}
\end{figure}
\begin{figure}
\begin{center}
{\epsfxsize=10 cm \epsfysize=8 cm {\epsfbox[31 330 589 716]
{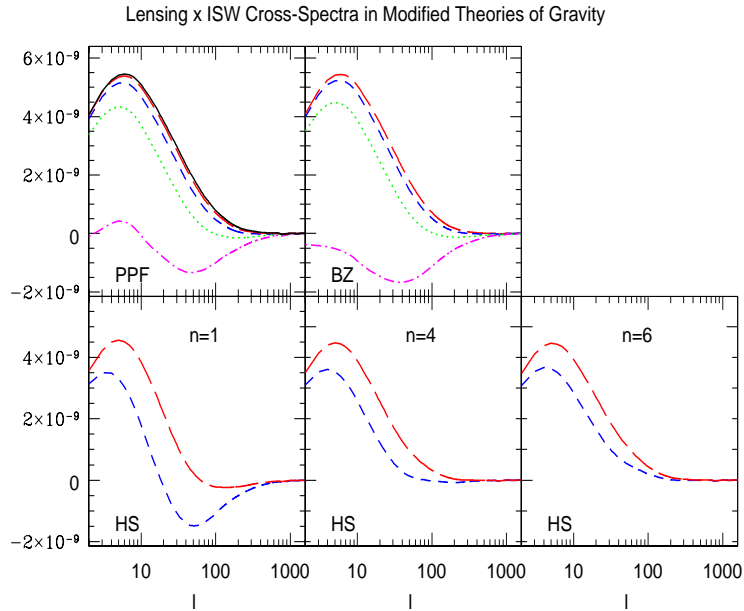}}}
\end{center}
\caption{Same as previous plot but for the ISW-lensing cross-spectra $\ell(\ell+1)\myC_{\ell}^{\phi{\rm T}}/2\pi$ defined 
in Eq.(\ref{eq:cross_spec_def}) as a function of the harmonic $\ell$ for various values of the parameter $\rB_0$.
The line-styles used for various models are same as that of Figure \ref{fig:clTT}.}
\label{fig:clphiT}
\end{figure}
We will be dealing with the secondary bispectra involving the
lensing of both primary anisotropies and other secondaries. Following
\citet{GoldbergSpergel99a}, \citet{GoldbergSpergel99b} and
\citet{CoorayHu} we start by expanding the observed temperature
anisotropy $\Theta(\oh)={\delta {\rm T}(\oh)/{\rm T}}$  in terms of the primary contribution $\Theta_{\rm
P}(\oh)$, the secondary contribution $\Theta_{\rm S}(\oh)$ and
lensing of the primary $\Theta_{\rm L}(\oh)$: 
\ben 
\Theta(\oh) =
\Theta_{\rm P} (\oh)+ \Theta_{\rm L} (\oh)+ \Theta_{\rm S}
(\oh)+\cdots. 
\een 
Here $\oh=(\theta,\phi)$ is the angular position
on the surface of the sky. Expanding the respective contributions in
terms of spherical harmonics $Y_{lm}(\oh)$ we can write: 
\ben
\Theta_{\mathrm {P}}(\oh) \equiv \sum_{\ell m} (\Theta_{\rm p})_{\ell m}
Y_{\ell m} (\oh); ~~~~ \Theta_{\mathrm {L}}(\oh) \equiv \sum_{\ell m}
[\nabla \psi(\oh) \cdot \nabla \Theta_{\mathrm {P}}(\oh)]_{\ell m}\;
Y_{\ell m}(\oh); ~~~~ \Theta_{\mathrm {S}}(\oh) \equiv \sum_{\ell m}
(\Theta_{\rm S})_{\ell m} Y_{\ell m} (\oh). 
\een 
Here $\psi(\oh)$ is the
projected lensing potential
\citep{GoldbergSpergel99a,GoldbergSpergel99b}. The secondary
bispectrum for the CMB takes contributions from products of P, L and
S terms with varying order. The bispectrum $B_{\ell_1\ell_2\ell_3}^{\rm PLS}$ is defined as follows (see \cite{Bartolo04} for generic discussion
of the bispectrum and its symmetry properties):
\begin{figure}
\begin{center}
{\epsfxsize=10 cm \epsfysize=8 cm {\epsfbox[28 324 594 726]
{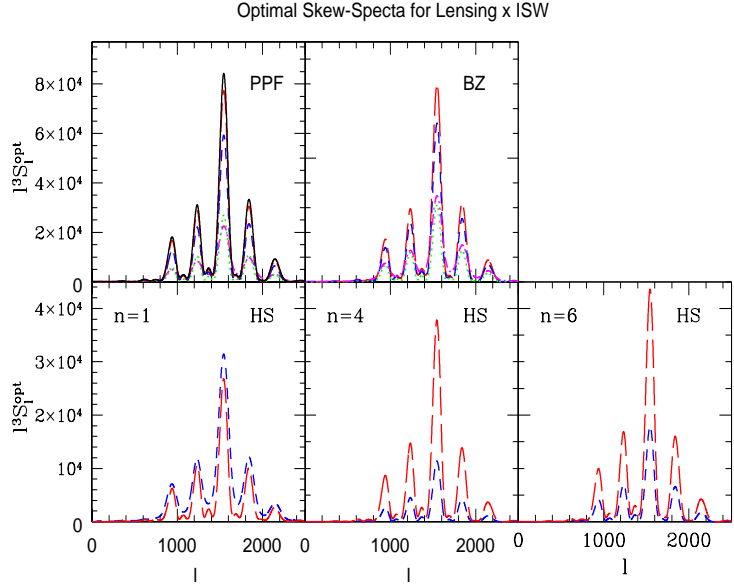}}}
\end{center}
\caption{The optimised skew-spectra  $\ell^3 S_{\ell}^{(\rm opt)}$, introduced in Eq.(\ref{eq:opt}) for various theories of modified gravity are displayed as a function
of harmonic $\ell$.  The general relativisitic (GR)
prediction corresponds to $\rB_0=0$ . The top-left and top-middle panels correspond to the predictions from PPF and 
BZ  respectively. The bottom panels correspond to HS with $n=1$ (bottom-left), $n=4$ (bottom-middle) and $n=6$ (bottom-right).
The values of ${\rm B_0}$ in these plots are ${\rm B}_0 = 10^{-2}$ and ${\rm B}_0 =10^{-3}$.
We have used $\ell_{\rm max}=2500$ and a Gaussian beam with FWHM $\theta_b=5'$ for the numerical evaluation of $S^{(\rm opt)}_{\ell}$. 
The line-styles used for various models are same as that of Figure \ref{fig:clTT}}
\label{fig:opt_hs}
\end{figure}
\begin{eqnarray}
B_{\ell_1\ell_2\ell_3}^{\rm PLS} && \equiv \sum_{m_1m_2m_3} \left ( \begin{array}{ c c c }
     \ell_1 & \ell_2 & \ell_3 \\
     m_1 & m_2 & m_3
  \end{array} \right) \int \left \langle \Theta_{\rm P}(\oh_1) \Theta_{\rm L}(\oh_2) \Theta_{\rm S}(\oh_3) \right \rangle
Y^*_{\ell_1m_1}(\oh_1) Y^*_{\ell_2m_2}(\oh_2) Y^*_{\ell_3m_3}(\oh_3) d \oh_1 d \oh_2 d \oh_3; \nonumber \\
&& \equiv \sum_{m_1m_2m_3} \left ( \begin{array}{ c c c }
     \ell_1 & \ell_2 & \ell_3 \\
     m_1 & m_2 & m_3
  \end{array} \right) \langle (\Theta_{\mathrm {\rm P}})_{\ell_1m_1} (\Theta_{\mathrm {\rm L}})_{\ell_2m_2}
(\Theta_{\mathrm {\rm S}})_{\ell_3m_3}  \rangle.
\label{eq:bispec_gen}
\end{eqnarray}
The angular brackets represent {\em ensemble} averages.
The matrices denote $3J$ symbols \citep{Ed68} and the asterisks denote complex conjugation.
It is possible to invert the relation assuming isotropy of the background Universe:
\ben
\langle (\Theta_{\mathrm {P}})_{\ell_1m_1} (\Theta_{\mathrm {L}})_{\ell_2m_2}
(\Theta_{\mathrm {S}})_{l_3m_3}  \rangle = \left ( \begin{array}{ c c c }
     \ell_1 & \ell_2 & \ell_3 \\
     m_1 & m_2 & m_3
  \end{array} \right) B^{\rm PLS}_{\ell_1\ell_2\ell_3}.
\een
Finally the bispectrum $B_{\ell_1\ell_2\ell_3}^{\rm PLS}$ is expressed in
terms of the unlensed primary power spectrum ${\cal
C}^{\rm TT}_\ell=\la(\Theta_{\rm P})_{lm}(\Theta_{\rm P}^*)_{\ell m}\ra$ and the
cross-spectra $\myC^{\phi T}_\ell$ (to be defined below) as follows: 
\ben 
&& B_{\ell_1\ell_2\ell_3}^{\rm PLS}  \equiv {b}^{\rm ISW-Len}_{\ell_1\ell_2\ell_3}I_{\ell_1\ell_2\ell_3}; \\
&& b^{\rm ISW-lens}_{\ell_1\ell_2\ell_3} = -{1 \over 2}\left [ \myC^{\phi {\rm T}}_{\ell_3} {\cal C}^{\rm TT}_{\ell_1}({\Pi_{\ell_2} - \Pi_{\ell_1} - \Pi_{\ell_3} })+ {\rm cyc.perm.}\right];\label{eq:define_R} \\
&& I_{\ell_1\ell_2\ell_3} \equiv\sqrt {\Xi_{\ell_1}\Xi_{\ell_2}\Xi_{\ell_3} \over 4\pi  }\left ( \begin{array}{ c c c }
    \ell_1 & \ell_2 & \ell_3 \\
     0 & 0 & 0
  \end{array} \right); \\ 
&& \Pi_{\ell} = \ell(\ell+1); \quad \Xi_{\ell} = (2\ell+1).
\label{eq:bispec_intro1} \een
See \citet{GoldbergSpergel99a},
\citet{GoldbergSpergel99b} for a derivation. The long-wavelength modes of ISW
couple with the short-wavelength modes of fluctuations 
generated due to lensing, hence the non-zero cross-spectrum $\myC^{\phi {\rm T}}_{\ell}$.
The reduced bispectrum
above is denoted as ${b}^{\rm ISW-Lens}_{\ell_1\ell_2\ell_3}$. To simplify the notation
for the rest of this paper, we henceforth drop the superscript $\rm
PLS$ from the bispectrum $B_{\ell_1\ell_2\ell_3}$. The cross-spectrum
$\myC^{\phi T}_{\ell}$ introduced above represents the cross-correlation
between the projected lensing potential $\psi(\oh)$ and the
secondary contribution $\Theta_{\rm S}(\oh)$: 
\ben
&& \la\psi(\oh)\Theta_{\rm S}(\oh')\ra = {1\over
4\pi}\sum_{\ell=2}^{\ell_{\rm max}}\Xi_{\ell}\,\myC^{\phi{\rm T}}_\ell
P_\ell(\hat\Omega\cdot\hat\Omega'), \label{eq:cross_spec_def} 
\een 
where $P_\ell$ are Legendre polynomials.
The cross-spectrum $\myC^{\phi T}_\ell$ takes different forms for ISW-lensing,
Rees-Sciama (RS)-lensing or Sunyaev-Zeldovich (SZ)-lensing correlations and we assume zero primordial
non-Gaussianity. The reduced bispectrum $b_{\ell_1\ell_2\ell_3}$
defined above using the notation $I_{\ell_1\ell_2\ell_3}$ is useful in
separating the angular dependence from the dependence on the power
spectra $\myC^{\phi {\rm T}}_\ell$ and $\myC^{\phi {\rm T}}_\ell$. We will use this to express the
topological properties of the CMB maps. The $\myC^{\phi {\rm T}}_\ell$ parameters for
lensing secondary correlations are displayed in
Figure \ref{fig:clphiT}. 

The beam $b_\ell(\theta_b)$ and the noise of a specific experiment are
characterised by the parameters $\sigma_{\rm beam}$ and $\sigma_{\rm
rms}$: 
\be 
b_\ell(\theta_b) = \exp[-\Pi\sigma_{\rm beam}^2]; \;\;\;
\sigma_{\rm beam} = {\theta_{b} \over \sqrt{8\ln(2)}}; \;\;\; n_\ell =
\sigma_{\rm rms}^2\Omega_{\rm pix}; \quad \Omega_{\rm pix} = {4\pi
\over {\rm N}_{\rm pix}}, \label{eq:beam_noise} 
\ee
where
$\sigma_{\rm rms}$ is the rms noise per pixel, that depends on the
full width at half maxima or FWHM of the beam, $\theta_{b}$. The
number of pixels ${\rm N}_{\rm pix}$ required to cover the sky
determines the size of the pixels $\Omega_{\rm pix}$. To incorporate
the effect of experimental noise and the beam we replace
${\cal C}_\ell \rightarrow {\cal C}_\ell b_\ell^2(\theta_b) + n_\ell$, and the
normalization of the skew-spectra that we will introduce later will
be affected by the experimental beam and noise. The computation of
the scatter will also depend on these parameters.

The reduced bispectrum for the unresolved point sources (PS) can be
characterized by a constant amplitude $b_{\rm PS}$ i.e. the angular 
averaged bispectrum $B^{\rm PS}_{\ell_1\ell_2\ell_3}$ for PS is given by $B^{\rm PS}_{\ell_1\ell_2\ell_3}=b_{\rm PS}I_{\ell_1\ell_2\ell_3}$;
for our numerical results we will take $b_{\rm PS}=10^{-29}$.

The optimal estimators for lensing-secondary mode-coupling
bispectrum have been recently discussed by \citep{MuVaCoHe11}. The
estimators that we propose here are relevant in the context of
constructing the MFs.
\begin{figure}
\begin{center}
{\epsfxsize=10 cm \epsfysize=7 cm {\epsfbox[31 319 586 716]
{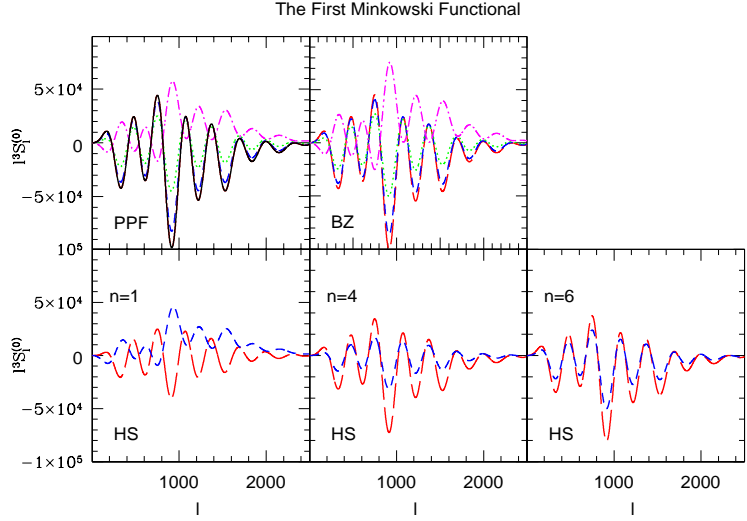}}}
\end{center}
\caption{The first skew-spectra associated with MFs, or the first Minkowski Spectra \, $\ell^3S_{\ell}^{(0)}$, defined in Eq.(\ref{sl1}), for various theories of modified gravity, displayed as a function of the harmonic $\ell$.  The top-left and top-middle panel correspond to PPF and BZ  respectively. The General Relativisitic (GR)
prediction corresponds to $\rB_0=0$  and is shown in the top-left panel (dot and long-dashed line). 
The bottom panels correspond to  HS for $n=1,4,6$ respectively.  The line-styles used for various models is same as Figure \ref{fig:clTT}.
}
\label{fig:sl0}
\end{figure}

\subsection{Computation of $\myC_\ell^{\rm TT}$, $\myC_\ell^{\phi \rm T}$ and $\myC_\ell^{\phi \phi}$}

The ISW effect and lensing potential $\phi$ can both be expressed in terms of the Weyl potential $\Phi -\Psi$:
\ben
&& {\delta T(\oh) \over T}\Big |_{\rm ISW} = \int dr {d \over d\tau} (\Phi -\Psi); \quad
\phi(\oh)  = -\int_0^{r_s} d r {r_s - r \over r r_s} (\Phi -\Psi).
\een
Assuming a flat Universe, we can express the cross-spectra $C_\ell^{\phi \rm T}$, the ISW contribution
$C_\ell^{\rm TT}$ to the power-spectrum, and the lensing potential spectrum  $C_\ell^{\phi \phi}$ as follows \citep{Hu2000}:
\ben
&& C^{\phi \rm T}_\ell=\frac{2\pi^2}{\ell^3}\int_0^{r_s} {\rm d}r \;r \; W^{\rm ISW}(r)W^{\rm Len}(r)\Delta^2_{\Phi}(k,0)\Big |_{k=\ell\frac{H_0}{r}} \;;\\
\label{eq:phiT}
&& C^{\rm TT}_\ell=\frac{2\pi^2}{\ell^3}\int_0^{r_s}  {\rm d}r \;r \; W^{\rm ISW}(r)W^{\rm ISW}(r)\Delta^2_{\Phi}(k,0)\Big |_{k=\ell\frac{H_0}{r}} \;;\label{eq:TT} \\
&& C^{\phi\phi}_\ell=\frac{2\pi^2}{\ell^3}\int_0^{r_s}  {\rm d}r \;r \; W^{\rm Len}(r)W^{\rm Len}(r)\Delta^2_{\Phi}(k,0)\Big |_{k=\ell\frac{H_0}{r}} \;,
\label{eq:phiphi}
\een
where $r_s$ is the comoving distance, and $r(z)=\int_0^z[{H_0}/{H(z')}]{\rm d}z'$.
\begin{figure}
\begin{center}
{\epsfxsize=10 cm \epsfysize=7 cm {\epsfbox[31 319 586 716]
{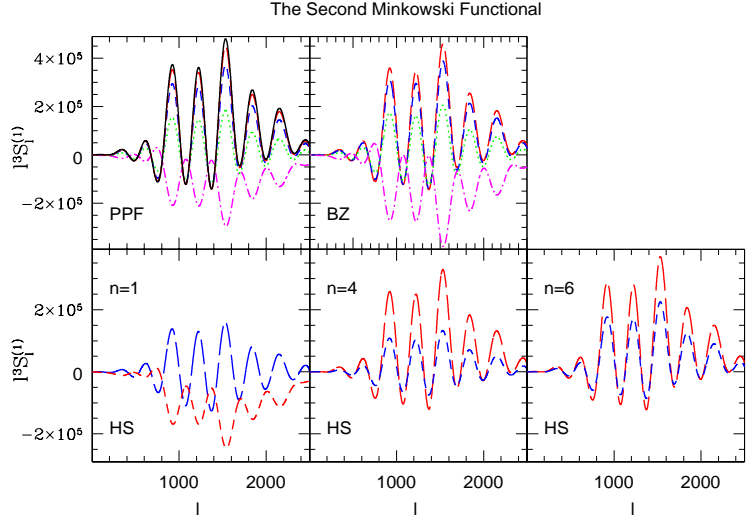}}}
\end{center}
\caption{Same as previous figure but for the second Minkowski Spectra, i.e. $\ell^3S^{(1)}_{\ell}$ defined in 
Eq.(\ref{sl2}). The line-styles used for various models is same as Figure \ref{fig:clTT}. }
\label{fig:sl1}
\end{figure}
\begin{figure}
\begin{center}
{\epsfxsize=10 cm \epsfysize=7 cm {\epsfbox[31 319 586 716]
{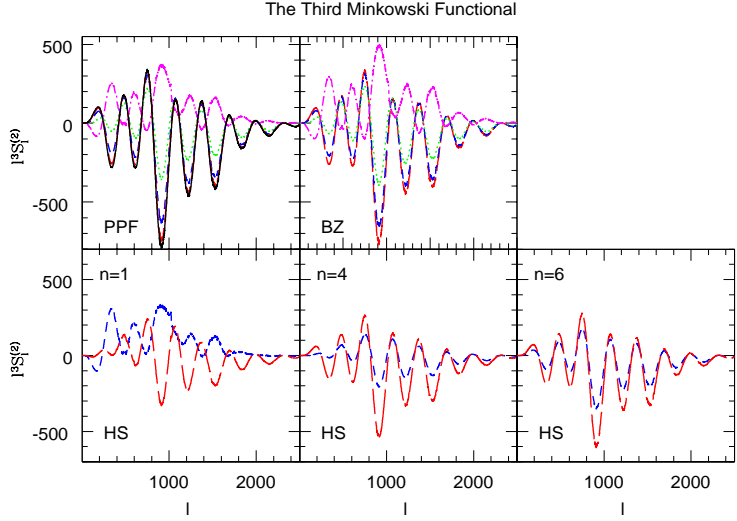}}}
\end{center}
\caption{Same as the previous figure but for the third Minkowski Spectra $\ell^3S^{(2)}_{\ell}$ defined in 
Eq.(\ref{sl3}). The line-styles used for various models are same as that of Figure \ref{fig:clTT}.}
\label{fig:sl2}
\end{figure}
We can express the gravitational potential power spectrum $\Delta^2_{\Phi}(k,z)$ by using the transfer function $T(k)$ and the growth factor $[F(z)/(1+z)]$:
\ben
&& \Delta^2_{\Phi}(k,z)=\frac{9}{4}\Omega_m^2\delta_H^2 F(z)T^2(k)\left(\frac{k}{H_0}\right)^{n-1}\;;
\een
with $\delta_H$ denoting the amplitude of matter density fluctuation at the present Hubble scale. The window functions
$W^{\rm ISW}(r)$ and $W^{\rm Len}(r)$ used above in Eq.(\ref{eq:phiT}) and Eq.(\ref{eq:TT}) are expressed as follows: 
\ben
&& W^{\rm{ISW}}(r)=-{{d} \over {d}r} [(1+\gamma){\rm{F}}]\;,\quad\quad W^{\rm{Len}}(r)=-(1+\gamma)F(r)\frac{(r_s-r)}{r\;r_s}\;.
\een
This is the expression used in Eq.(\ref{eq:bispec_intro1}) to construct the bispectrum
which was used to compute the optimised skew-spectra of Eq.(\ref{eq:opt}) and the sub-optimal versions in Eq.(\ref{sl1})-Eq.(\ref{sl3}) to be introduced 
in \textsection\ref{sec:MF} later.
\section{Minkowski Functionals and associated Power-Spectra}
\label{sec:MF}
The study of non-Gaussianity is usually primarily focused on the
bispectrum, as this saturates the Cram\'er-Rao bound
\citep{Babich,KSH11} and is therefore in a sense optimal. However in
practice it is difficult to probe the entire configuration
dependence using
noisy data \citep{MuHe10}. An alternative is to use cumulant correlators, which are multi-point
correlators collapsed to encode two-point statistics. These were
introduced into galaxy clustering by
\cite{SS99}, and were later found to be useful for analyzing
projected surveys such as the APM galaxy survey \citep{MuMeCo00}. Being
two-point statistics they can be analyzed in multipole space by
defining an associated power spectrum \citep{Cooray01}. Recent studies 
by \cite{CLM08} have demonstrated their wider
applicability including, e.g., in 21cm studies. In more recent
studies the skew- and kurt-spectra were found to be useful for
analysing temperature \citep{MuHe10} as well as polarization maps \citep{MuCoCooHeSm11} and from maps of secondaries
from CMB experiments \citep{MuSmJouCo,MuCoHe13} and in weak lensing studies
\citep{MuWaSmCo12}.
The MFs are well known morphological descriptors which are used in the study of random fields. 
Morphological properties are the properties that remain invariant under rotation and translation (see \cite{Hadwiger59}
for a more formal introduction). They are defined
over an excursion set $\Sigma$ for a given threshold $\nu$. The three MFs that are defined for two dimensional (2D)
studies can be expressed as \citep{PratMun12}:
\be
V_0(\nu) = \int_{\Sigma} da; \quad V_1(\nu) = {1 \over 4}\int_{\partial\Sigma} dl; \quad V_2(\nu) = {1 \over 2\pi}\int_{\partial \Sigma}\kappa dl
\ee
Here $da$, $dl$ are the elements for the excursion set $\Sigma$ and its boundary $\partial \Sigma$. The MFs $V_k(\nu)$
correspond respectively to the area of the excursion set $\Sigma$, the length of its boundary $\partial\Sigma$, and the
integral curvature along its boundary (which is also related to the genus $g$ and hence the Euler characteristics $\chi$).

Following earlier studies \citep{MuSmJoCo13,MuWaSmCo12,MuCoHe13} we introduce 
three different skew-spectra associated with MFs for an arbitrary cosmological projected field $\Psi$:
\ben
&& S_\ell^{(0)} \equiv {1 \over N_{0}}S_\ell^{(\myf^2,\myf)} \equiv {1 \over N_0}{1 \over \Xi_{\ell}}\sum_m 
{\rm Real}([\myf]_{\ell m}[\myf^2]^*_{\ell m})  ={1 \over N_0} \sum_{\ell_1 \ell_2} B_{\ell\ell_1l_2}J_{\ell\ell_1\ell_2}
\label{sl1} \\
&& S_\ell^{(1)} \equiv {1 \over N_1}S_\ell^{(\myf^2,\nabla \myf)} 
\equiv {1 \over N_1}{1 \over \Xi_{\ell}}\sum_m
{\rm Real}([\nabla^2 \myf]_{\ell m}[\myf^2]^*_{\ell m}) 
={1 \over N_1} \sum_{\ell_1\ell_2} \Big [{\Pi_{\ell}+ \Pi_{\ell_1}+ \Pi_{\ell_2}} \Big ] B_{\ell\ell_1\ell_2}J_{\ell\ell_1\ell_2}
\label{sl2} \\
&& 
S_\ell^{(2)} \equiv {1 \over N_2}S_\ell^{(\nabla \myf\cdot\nabla \myf, \nabla^2\myf)} \equiv 
{1 \over N_2}{1 \over \Xi_{\ell}}\sum_m
{\rm Real}([\nabla \myf \cdot \nabla \myf]_{\ell m}[\nabla^2 \myf]^*_{lm})\nn \\
&&\quad\quad\quad ={1 \over N_2} \sum_{\ell_1\ell_2}
{1 \over 2}\Big [ [\Pi_{\ell}+\Pi_{\ell_1} - \Pi_{\ell_2}]\Pi_{\ell_2} + {\rm cyc.perm.} \Big ]
 B_{\ell\ell_1\ell_2}J_{\ell\ell_1\ell_2} \label{sl3}\\
&& J_{\ell_1\ell_2\ell_3} \equiv {I_{\ell_1\ell_2\ell_3} \over \Xi_{\ell_1}} = \sqrt{\Sigma_{\ell_2}\Sigma_{\ell_3} \over \Sigma_{\ell_1} 4 \pi }\left ( \begin{array}{ c c c }
     \ell_1 & \ell_2 & \ell_3 \\
     0 & 0 & 0
  \end{array} \right); \label{eq:defineJ3}\\
&& S^{(i)} = \sum_{\ell}\Xi_{\ell}S^{(i)}_{\ell};\\ 
&& N_0 = 12\pi\sigma_0^4; \quad  N_1 = 16\pi\sigma_0^2\sigma_1^2; \quad N_2 = 8\pi\sigma_1^4.
\label{eq:S_l}
\een
\begin{figure}
     \begin{center}
        \subfigure[]{%
            \label{fig:first}
            \includegraphics[width=0.32\textwidth]{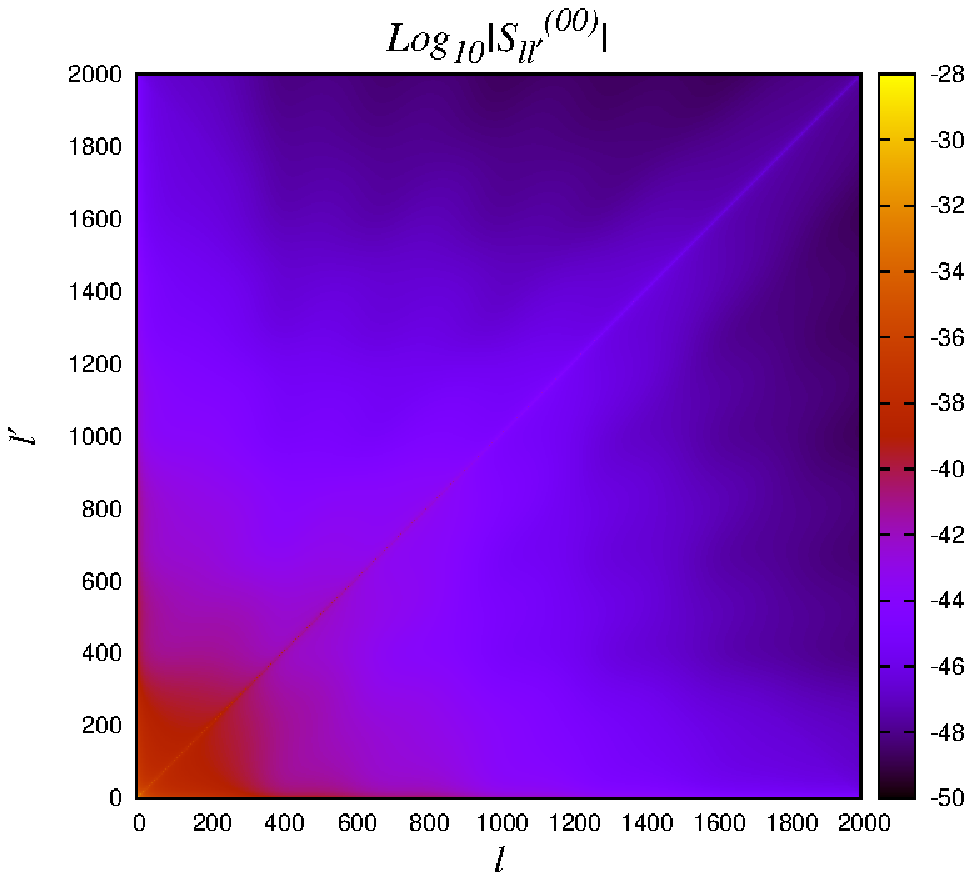}
        }%
        \subfigure[]{%
           \label{fig:second}
           \includegraphics[width=0.32\textwidth]{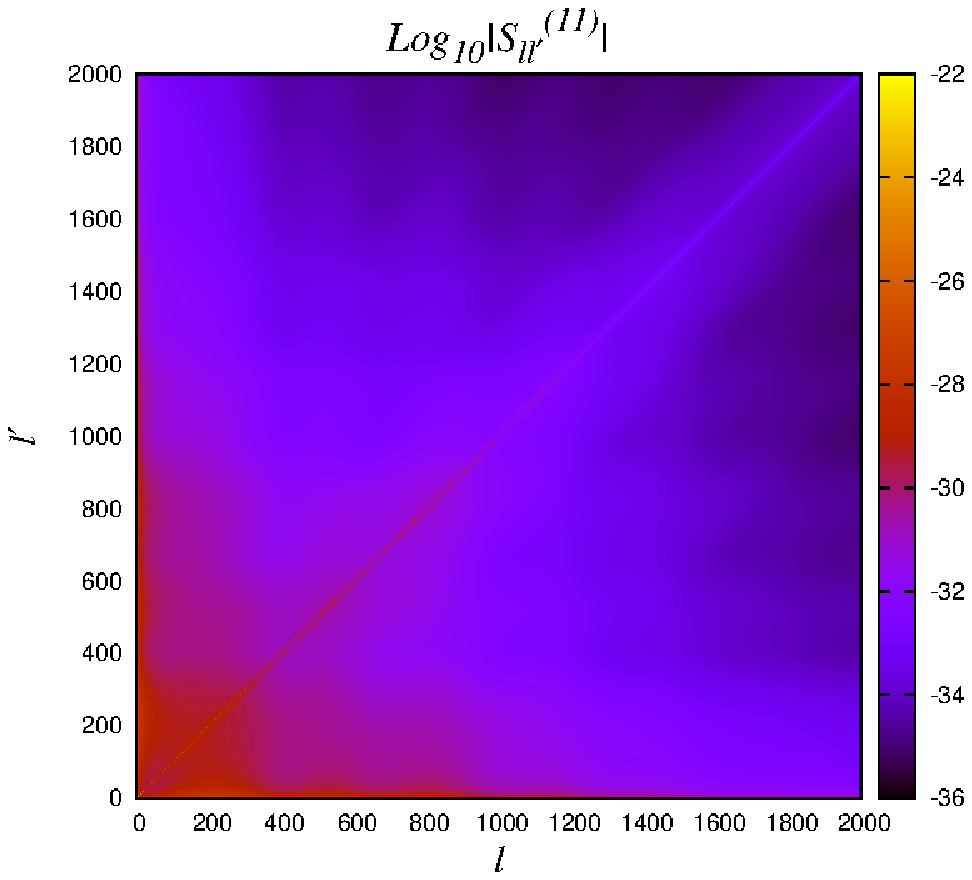}
        }
        \subfigure[]{%
            \label{fig:third}
            \includegraphics[width=0.32\textwidth]{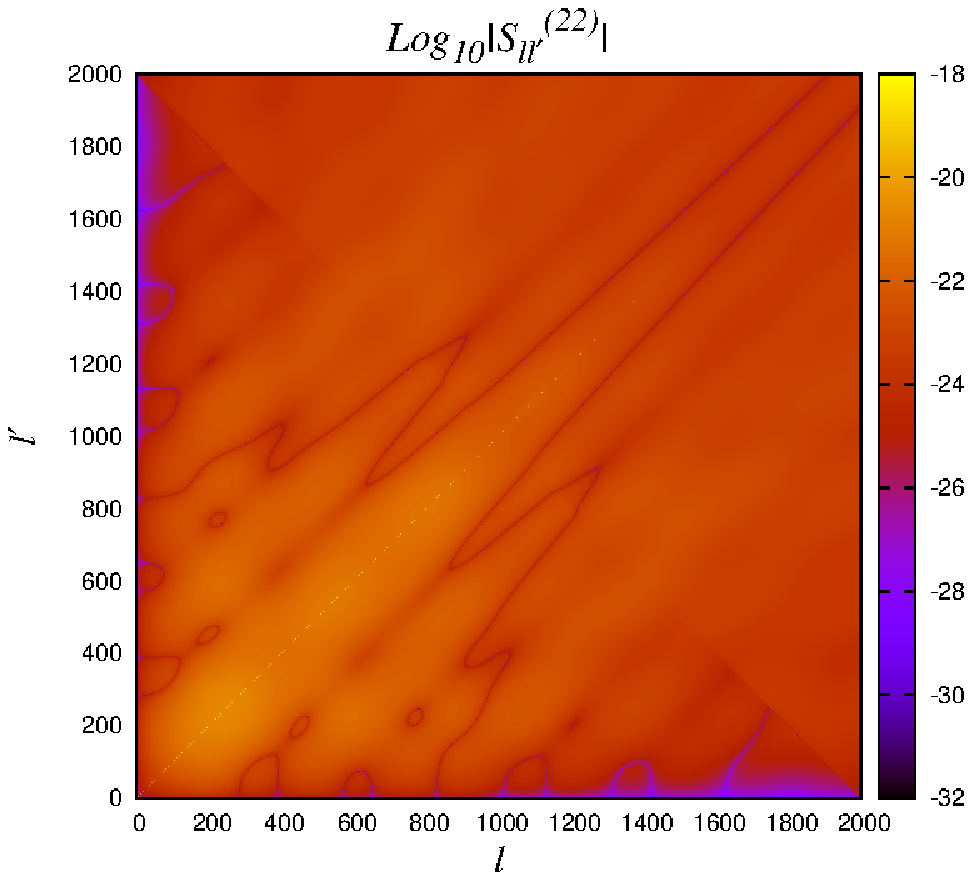}
        }%
%
    \end{center}
    \caption{%
      The all-sky covariance matrix $S^{(00)}_{\ell\ell'}$ for the estimator $S^{(0)}_{\ell}$ is being plotted in the left panel.
      In middle panel we depict  $S^{(11)}_{\ell\ell'}$ and the right panel correspond to $S^{(22)}_{\ell\ell'}$.
      The analytical expressions for the covariance matrices are given in Eq.(\ref{eq:covSij}).
      For the computation of these covariance matrices we assume $\rB_0=0$ (GR).
     }%
   \label{fig:subfigures}
\end{figure}
In contrast to these MF-based quantities, the optimised skew-spectra $S_\ell^{\rm opt}$ for two different types of non-Gaussianity is defined by the following expression:
\ben
S_\ell^{\rm opt}(\rm{X,Y}) = 
{1 \over 6}\sum_{\ell_1\ell_2} {B^{\rm X}_{\ell\ell_1\ell_2}B^{\rm Y}_{\ell\ell_1\ell_2} \over \myC^{\rm tot}_{\ell}\myC^{\rm tot}_{\ell_1}\myC^{\rm tot}_{\ell_2}}; 
\quad\quad S^{\rm opt}(\rm{X,Y}) = \sum_{\ell} S_\ell^{\rm opt}(\rm{X,Y}).
\label{eq:opt}
\een
The three skew-spectra associated with MFs, defined in Eq.(\ref{sl1})-Eq(\ref{sl3}), for various theories of modified gravity are shown in Figure \ref{fig:sl0} as a function of  harmonic $\ell$ for PPF, BZ and HS models. 
Clearly the one-point estimator defined in Eq.(\ref{eq:S_l})
will have nearly vanishing amplitude due to cancellation originating from the oscillatory pattern seen in all three skew-spectra associated with MFs.
The FWHM is fixed at $\theta_b=5'$. The noise level is chosen to match the Planck 143GHz channel. It is interesting to note
that the extrema of $\ell^3S_{\ell}^{(0)}$ for all models occurs roughly at similar $\ell$ values.
We display four different values of $\rB_0$ for each models $\rB_0=10^{-3}$ (solid), $\rB_0=10^{-2}$ (short-dashed), $\rB_0=10^{-1}$ (long-dashed) and $\rB_0=1$ (dot-dashed) respectively.
For HS models we choose two different values for $\rB_0$ i.e. $\rB_0=10^{-3}$ and $\rB_0=10^{-2}$. In agreement with what we found
for optimised estimators the skew-spectra for the HS models with low $n$ values show a greater degree of sensitivity to $\rB_0$
compared to their high-$n$ counterparts, which roughly mimic their PPF or BZ counterparts.
\section{Likelihood Analysis Using Skew-Spectra}
\label{sec:like}
In this section we construct the joint covariance matrices for skew-spectra and ordinary spectra and provide
results of a likelihood analysis forecast for the parameter $\rm B_0$. 

The Gaussian contributions to the covariance matrix can be expressed in terms of the total power-spectrum $\myC^{\rm tot}_{\ell}$ alone;
which in terms of beam $b_\ell$ and the noise power spectrum $n_\ell$ takes the form 
$\myC^{\rm tot}_{\ell}=\myC^{\rm TT}_{\ell}b_{\ell}^2+n_{\ell} $:
\ben
&& {\cal S}_{\ell\ell'}^{(ij)} = \la \delta S_{\ell}^{(i)} \delta S^{(j)}_{\ell'}\ra= \la S_\ell^{(i)}S_{\ell'}^{(j)} \ra - \la S_\ell^{(i)}\ra \la S_{\ell'}^{(j)}\ra \nn \\
&&\quad\quad\quad\quad = {1 \over N_{(i)}} { 1 \over N_{(j)}}\myC^{\rm tot}_{\ell}\sum_{\ell_1\ell_2}\; \myC^{\rm tot}_{\ell_1}\myC^{\rm tot}_{\ell_2}
J^{(i)}_{\ell\ell_1\ell_2}\Big [\left ( J^{(j)}_{\ell'\ell_1\ell_2} +J^{(j)}_{\ell'\ell_2\ell}+J^{(j)}_{\ell'\ell\ell_1} \right )
+ (-1)^{\ell+\ell_1+\ell_2}\left (  J^{(j)}_{\ell'\ell_2\ell_1} +J^{(j)}_{\ell'\ell\ell_2}+J^{(j)}_{\ell'\ell_1\ell}  \right ) 
\Big ];\quad\quad \label{eq:cov1}\\
&& J_{\ell_1\ell_2\ell_3}^{(0)}= J_{\ell_1\ell_2\ell_3}; \quad
J_{\ell_1\ell_2\ell_3}^{(1)}= \left ( \Pi_{\ell_1}+\Pi_{\ell_2}+\Pi_{\ell_3} \right )J^{}_{\ell_1\ell_2\ell_3};\quad
J_{\ell_1\ell_2\ell_3}^{(2)}= \left ( (\Pi_{\ell_1}+\Pi_{\ell_2}-\Pi_{\ell_3})\Pi_{\ell_3} +\rm{cyc.perm.} \right )J^{}_{\ell_1\ell_2\ell_3}.
\een
We use the following expression in our derivation \citep{Bartolo04}:
\ben
&& \langle B_{\ell_1\ell_2\ell_3}B_{\ell'_1\ell'_2\ell'_3} \rangle = \myC^{\rm tot}_{\ell_1}\myC^{\rm tot}_{\ell_2}\myC^{\rm tot}_{\ell_3}
 \Big [\Big (\delta^{\ell_1'\ell_2'\ell_3'}_{\ell_1\ell_2\ell_3}+\delta^{\ell_3'\ell_1'\ell_2'}_{\ell_1\ell_2\ell_3}
+ \delta^{\ell_2'\ell_3'\ell_1'}_{\ell_1\ell_2\ell_3} + (-1)^{\ell_1+\ell_2+\ell_3} \left(\delta^{\ell_1'\ell_3'\ell_2'}_{\ell_1\ell_2\ell_3}+\delta^{\ell_2'\ell_1'\ell_3'}_{\ell_1\ell_2\ell_3}
+ \delta^{\ell_3'\ell_2'\ell_1'}_{\ell_1\ell_2\ell_3}\right ) \Big ]; \nn \\ 
&& \delta^{\ell_1'\ell_2'\ell_3'}_{\ell_1\ell_2\ell_3} = \delta_{\ell_1\ell_1'}\delta_{\ell_2\ell_2'}\delta_{\ell_3\ell_3'}.
\een
Notice that the $3\rm J$
symbols involved in the definitions of $S^{(i)}_{\ell}$ all have the azimuthal quantum numbers $m_i=0$ in which case 
we have non-zero $3\rm J$ symbols only when $(\ell+\ell_1+\ell_2) ={\rm even}$, thus $(-1)^{\ell+\ell_1+\ell_2}=1$. Thus
we notice that $J^{(i)}_{\ell\ell_1\ell_2}$ is symmetric under the exchange of the last two indices i.e. $\ell_1$ and $\ell_2$.
Using these facts, after a straightforward but tedious calculation, we can further simplify Eq.(\ref{eq:cov1}) to \,:
\ben
&& {\mathbb S}^{(ij)}_{\ell\ell'} \equiv {\cal S}_{\ell\ell'}^{(ij)} = \la \delta S_{\ell}^{(i)} \delta S^{(j)}_{\ell'}\ra= {1 \over N_{(i)}} { 1 \over N_{(j)}}
\Big [2\; \delta_{\ell\ell'}\; \myC^{\rm tot}_{\ell}\sum_{\ell_1\ell_2}\; \myC^{\rm tot}_{\ell_1}\myC^{\rm tot}_{\ell_2}
J^{(i)}_{\ell\ell_1\ell_2}J^{(j)}_{\ell\ell_1\ell_2}  
+ 4\; \myC^{\rm tot}_{\ell} \myC^{\rm tot}_{\ell'}\sum_{\ell_1}\myC_{\ell_1}\; J^{(i)}_{\ell\ell'\ell_1}J^{(j)}_{\ell'\ell\ell_1} \Big ].
\label{eq:covSij}
\een
The first term contributes only to diagonal entries of the covariance matrix while the second term contributes also to
the off-diagonal terms. This is the expression we have used in our numerical computations. The covariance matrix involving the bispectrum derived above is generic but depends on the assumption that the non-Gaussianity is weak i.e. $\la B_{\ell_1\ell_2\ell_3} \ra\simeq 0$ and can also be used for 
likelihood calculations of primordial non-Gaussianity using MFs \,\citep{MuSmCooReHeCo13}.

For the one-point estimators introduced previously,  $S^{(i)} = \sum_{\ell}\Xi_{\ell}S^{(i)}_{\ell}$ the covariance matrix ${\cal S}^{(ij)}$ takes the following form:
\ben
&&  {\cal S}^{(ij)} = \la \delta S^{(i)} \delta S^{(j)}\ra= \sum_{\ell\ell'} \Xi_{\ell}\Xi_{\ell'}{\cal S}_{\ell\ell'}^{(ij)} = {1 \over N_{(i)}} { 1 \over N_{(j)}}\sum_{\ell_1\ge\ell_2\ge\ell_3}  \myC^{\rm tot}_{\ell_1}\myC^{\rm tot}_{\ell_2} \myC^{\rm tot}_{\ell_3} \; I^{(i)}_{\ell_1\ell_2\ell_3}I^{(j)}_{\ell_1\ell_2\ell_3} .
\een
\begin{figure}
\begin{center}
{\epsfxsize=12 cm \epsfysize=6 cm {\epsfbox[31 514 586 716]
{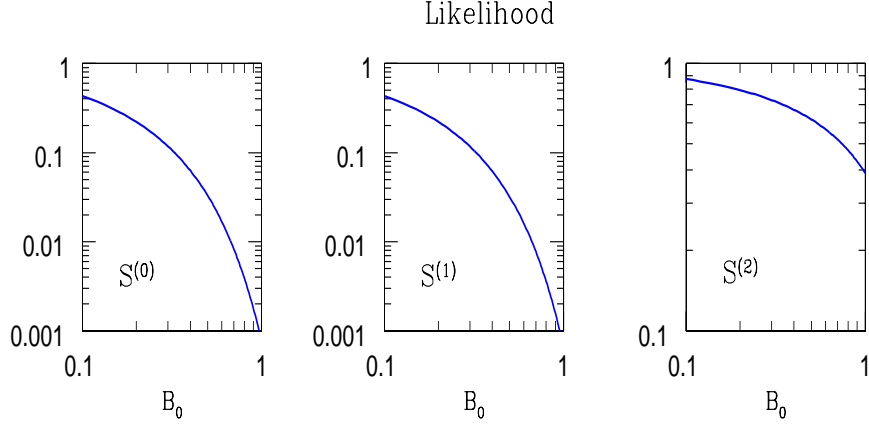}}}
\end{center}
\caption{The likelihood function ${\cal L}_{\rm S}$ defined in Eq.(\ref{eq:like1}) for estimators $S_\ell^{(0)}$ (left-panel), $S_\ell^{(1)}$ (middle-panel) and $S_\ell^{(2)}$ (right-panel) are plotted as a function of $\rB_0$. The parametrization used in our computation is that of BZ;
primarily due to it's higher speed compared to other parametrisation in numerical implementation.}
\label{fig:like}
\end{figure}

Finally, the covariance of the optimum estimator $S^{\rm opt}_{\ell}$  defined in Eq.(\ref{eq:opt}) 
is given by the following expression:
\ben
&& \la\delta S_{\ell}^{\rm opt}\delta S_{\ell'}^{\rm opt}\ra = 
{1 \over 18} \delta_{\ell\ell'} \sum_{\ell_a\ell_b}{B^2_{\ell\ell_a\ell_b} \over \myC^{\rm tot}_{\ell}
\myC^{\rm tot}_{\ell_a}\myC^{\rm tot}_{\ell_b}}
+ {1 \over 9}\sum_{l_a}{B^2_{\ell\ell'\ell_a} \over \myC^{\rm tot}_{\ell}\myC^{\rm tot}_{\ell'}\myC^{\rm tot}_{\ell_a}}
  = {1 \over 3} \delta_{\ell\ell'}S^{\rm opt}_{\ell}+ {1 \over 9} \sum_{l_a}{B^2_{\ell\ell'\ell_a} \over \myC^{\rm tot}_{\ell}\myC^{\rm tot}_{\ell'}\myC^{\rm tot}_{\ell_a}}.\label{eq:covopt}\\
&& {\mathbb S}^{\rm opt} \equiv \la\delta S^{\rm opt}\delta S^{\rm opt}\ra = \sum_{\ell} S^{\rm opt}_{\ell} = S^{\rm opt}.
\een
This result agrees with the previous calculation of \cite{MuHe10}, using Fisher matrices in the limit 
of all-sky coverage. The results given there include additional correction terms (termed ``$\beta$''), related to the so called {\em linear}, and {\em cubic} (``$\alpha$'' ) terms, due to partial sky coverage.  
The likelihoods for the MFs and the optimal skew-Cls are
\ben
&& {\cal{L}}_{\rm S} = \exp(-\chi_{\rm S}^2/2); \quad\quad  \chi^2_{\rm S} = \sum_{ij}\sum_{\ell\ell'}\left [ \delta S^{(i)}_{\ell} [{\mathbb S^{(ij)}}]^{-1}_{\ell\ell'}  \delta S^{(i)}_{\ell'}\right ]; \label{eq:like1}\\
&& {\cal{L}}_{\rm opt} = \exp(-\chi_{\rm opt}^2/2); \quad\quad 
\chi^2_{\rm S} = \sum_{\ell\ell'}\left [ \delta S^{\rm opt}_{\ell} [{\mathbb S^{\rm opt}}]^{-1}_{\ell\ell'}  \delta S^{\rm opt}_{\ell'}\right ]. \label{eq:like2}
\een
For corresponding one-point estimators we have ${\cal{L}}_{\rm S} = \exp(-{[\delta S^{\rm opt}]^2/2\mathbb S})$ and similarly for joint
analysis using all one-point MFs  ${\cal{L}}_{\rm S} = \exp(-[\delta S^{(i)}][{\mathbb S^{(ij)}}]^{-1}[\delta S^{(j)}]/2)$.

{\bf Bayesian Recovery of ${\rm B}_0$:} In recent works, \cite{Hikage08} and \cite{Du13} adopted a Bayesian approach in
their analysis of primordial non-Gaussianity in CMB maps using MFs.
We can similarly use Bayes' theorem to write the posterior probability for 
$\rm B_0$,  ${\rm P}({\rm B}_0|{\bf S})$ given the one-point MFs as the data vector ${\bf S}$:
\ben
&& {\rm P}({\rm B}_0|{\bf S}) ={ {\cal L}_{\rm S}({\bf S} | {\rm B}_0) {\rm P}({\rm B}_0) \over \int {\cal L}_{\rm S}({\bf V}|{\rm B}_0){\rm P}({\rm B}_0)d{\rm B}_0}; \quad
{\bf S} = (S^{(0)},S^{(1)},S^{(2)}).
\label{eq:Bayes}
\een
Here ${\rm P}({\rm B}_0)$ is the prior, assumed flat. Similarly we can also use the optimized skewness as the data vector instead of the MFs by replacing 
${\bf S}$ by $S^{\rm opt}$ and the likelihood function by  ${\cal L}_{\rm opt}({\bf S} | {\rm B}_0)$.
The likelihood function in such studies is typically assumed to be Gaussian, or determined using Monte Carlo simulations.
We find that the likelihood for $\rm B_0$ has an extended non-Gaussian tail. Thus, the analytical covariance and the
corresponding likelihood derived here will be useful in providing independent estimates, and related error-bars for sanity checks
of results derived through Monte-Carlo simulations.

\begin{figure}
\begin{center}
{\epsfxsize=8 cm \epsfysize=7 cm {\epsfbox[30 16 329 263]
{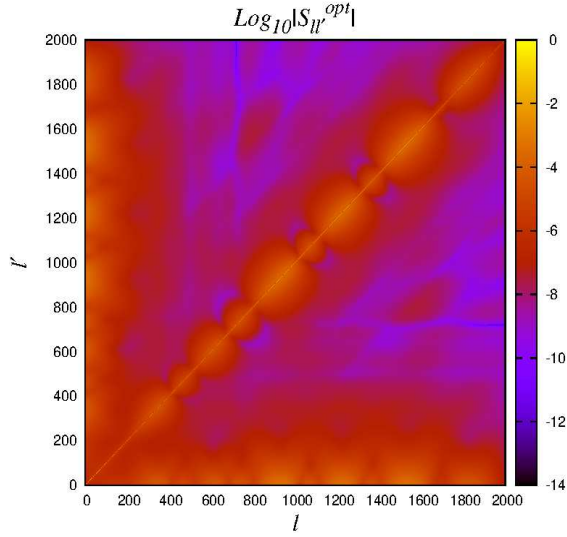}}}
\end{center}
\caption{The all-sky covariance matrix ${\mathbb S}^{\rm opt}_{\ell\ell'}$ defined in Eq.(\ref{eq:covopt}) for the 
optimal estimator $S^{\rm opt}_{\ell}$ is shown. The experimental set up corresponds to Planck 143 GHz channel with $\ell_{max}=2000$.
The mode-mode coupling, even in the case of all-sky coverage, 
seen here in the covariance matrix is a result of the facet that skew-spectrum is a non-Gaussian
statistics. In the notation of \citep{MuHe10} the covariance matrix presented here comprise of only the ``$\alpha$'' terms.
Additional mode-coupling is expected as pointed out in \citep{MuHe10}. The resulting ``$\beta$'' terms 
are subdominant for near all-sky coverage. We assumed a homogeneous uncorrelated noise distribution in our calculation, 
see \citep{MuHe10} for a complete treatment. We also assumed $\rB_0=0$ (GR) background for our computation.}
\label{fig:covmat}
\end{figure}
\section{Results}
\label{sec:result}
We have introduced three different MFs in this study and compared their performance against the optimum estimator.
The aim is to use CMB data to constrain the departure of modified gravity theories from GR as parametrized by the parameter $B_0$
that denotes the Compton wavelength of the scalaron at the present epoch. The underlying bispectrum that we probe is the one
generated by correlation between ISW and lensing of the CMB. The bispectrum is constructed from 
$\myC^{\rm TT}_{\ell}$ (Figure \ref{fig:clTT}) and $\myC^{T\phi}_{\ell}$ (Figure \ref{fig:clphiT}). 

The set three skew-spectra associated with MFs or the first Minkowski Spectra \, $\ell^3S_{\ell}^{(0)}$, 
defined in Eq.(\ref{sl1})-Eq.(\ref{sl3}), for various theories of modified gravity are displayed in 
Figure \ref{fig:sl0} - Figure \ref{fig:sl2} as a function of the harmonic $\ell$.  The top-left and top-middle panels in these figures
corresponds to predictions from PPF and BZ respectively. The General Relativistic (GR)
prediction correspond to $\rB_0=0$  and is shown in the top-left panel (dot and long-dashed line). 
The bottom panels correspond to the results from the HS model, for $n=1,4,6$ respectively. It is interesting to note that
the the one-point estimator defined in Eq.(\ref{eq:S_l})
will have nearly vanishing amplitude due to cancellation originating from the oscillatory pattern seen in all three skew-spectra associated with MFs - which is one of the motivation for studying the associated power-spectra.
The FWHM is fixed at $\theta_b=5'$. The noise level is chosen to match the Planck 143GHz channel. It is interesting to note
that the extrema of $\ell^3S_{\ell}^{(0)}$ for all models occurs roughly at similar $\ell$ values and thus are relatively insensitive to
the change in parameter $B_0$. We display four different values of $\rB_0$ for each model: $\rB_0=10^{-3}$ (solid), $\rB_0=10^{-2}$ (short-dashed), $\rB_0=10^{-1}$ (long-dashed) and $\rB_0=1$ (dot-dashed) respectively.
For HS models we choose two different values for $\rB_0$ i.e. $\rB_0=10^{-3}$ and $\rB_0=10^{-2}$. In agreement with what we found
for optimised estimators the skew-spectra for HS models with low $n$ values show greater degree of sensitivity to $\rB_0$
compared to their higher $n$ counterparts, that roughly mimic their PPF or BZ counterparts. The corresponding optimum spectrum
is given in Figure \ref{fig:opt_hs}. By construction the optimum skew-spectra are positive definite. The peak structure of
the optimum estimator for a given model is different from its MFs counterparts. The odd-numbered peaks of the optimum estimator
are much more pronounced compared to their even-numbered counterparts. Increasing the value of $B_0$ suppresses the amplitude of
oscillations for both Minkowski Spectra and the optimum skew-spectra.

We have derived the covariance of the Minkowski Spectra and optimum skew-spectra. The covariance of Minkowski Spectra 
depends only on the ordinary temperature power spectrum and are independent of the bispectrum as they are
derived in the limiting case of vanishing bispectrum.  The covariance of the optimum skew-spectrum 
depends on the target bispectrum used for the construction of weights. Both set of covariance matrices are well-conditioned. 
The analytical covariance matrices were derived using an all-sky approximation. The mode-mode coupling despite
the all-sky approximation is related to the fact that these statistics are inherently non-Gaussian. 
The covariance matrices for the MFs are displayed in Figure \ref{fig:subfigures} and for the optimum skew-spectra they are displayed in
Figure \ref{fig:covmat}. A comparison with results presented 
in \cite{MuHe10} shows that we recover the terms designated as ``$\alpha$'' term there. The lack of corresponding ``$\beta$''
terms in the current study is simply due to all-sky coverage assumed here for simplicity.  
%
%
\begin{figure}
\begin{center}
{\epsfxsize=5 cm \epsfysize=5 cm {\epsfbox[21 430 309 716]
{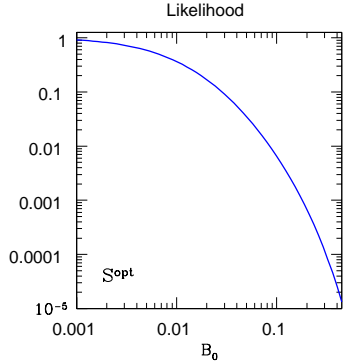}}}
\end{center}
\caption{Likelihood for optimal-estimator $S_{\ell}^{\rm opt}$ is plotted as a function of ${\rm B}_0$ using the
covariance matrix ${\mathbb S}^{\rm opt}_{\ell\ell'}$ defined in Eq.(\ref{eq:covopt}).
The parametrization used in this computation is that of BZ.}
\label{fig:opt_like}
\end{figure}

Finally we use these covariance matrices to compute the
likelihood functions. 
\cb{The results are obtained by using a fiducial value $\rm B_0=0.$ The {\em analytical} covariance matrix for the 
optimum estimators are described 
in Eq.(\ref{eq:opt}). These expression was used in association with  Eq.(\ref{eq:like2})
to compute the likelihood function presented in Figure \ref{fig:opt_like}.
The likelihood functions of $\rB_0$ for MFs are shown in Figure \ref{fig:like}. In this case,
we use the Eq.(\ref{eq:covSij}) for the expression of covariance matrices and in Eq.(\ref{eq:like1})
for the expression of likelihood function. We find $B_0< 0.67$ and $B_0<0.45$ for both $S^{(0)}_{\ell}$ 
and $S^{(1)}_{\ell}$ at $99\%$ and $95\%$ confidence level respectively. For $S^{(opt)}_{\ell}$ 
the numbers are $0.071$($99\%$ CL) and $0.15$ ($95\%$ CL) respectively. It's important to realise that
the liklihood functions are {\em not} Gaussian as the covariance matrices
depend on $B_0$ in a non-trivial manner through their dependence on the CMB temperature power spectrum $\myC^{\rm TT}_{\ell}$
and the lensing-temperature cross-spectrum $\myC^{\phi\rm T}_{\ell}$. }


\section{Discussion and Conclusions}
\label{sec:disc}
The correlation between ISW and lensing of the CMB generates a specific signature in the CMB bispectrum.
Analysis of first-year data from the Planck satellite has detected this signature with a moderate level of signal to noise ($2.6\sigma$).
The ISW-lensing bispectrum is unique as it depends on the CMB power-spectrum generated at recombination
and cross-spectra of the lensing potential and the ISW effect generated at late times. Both ISW and lensing
are sensitive to the underlying model of gravity, and thus the resulting bispectrum
provides an opportunity to constrain any departure from GR.
We consider various formulations of the modified gravity models which include HS, BZ and PPF 
models to compute the bispectrum.

{\bf Topological Estimators: }The non-Gaussianity in CMB maps are often studied using moment-based approaches
or alternatively using their harmonic counterparts, namely the multi-spectra.
Extending previous results we have studied how topological descriptors such as the MFs
can provide a complementary role, paying special attention to Planck-type experiments.
The MFs are interesting as they have different responses to various systematics.
We have considered the three MFs that are used for describing the topological
properties of CMB temperature maps. We compute analytically the covariance associated with the
skew-spectra associated to the MFs, and our results also include cross-covariance among different
skew-spectra. In agreement with previous results we find that the skew-spectra are
highly correlated. Constructing the MFs for Planck type experiments (143 GHz). We find that the constraints are tighter for 
the first two MFs $S_{\ell}^{(0)}$ and 
$S_{\ell}^{(1)}$, which both give ${\rm B}_0<0.67$ at $95\%$ CL. We do not get any meaningful constraints using $S^{(2)}_{\ell}$.  
The constraints can be further improved by considering Wiener filtering instead of
Gaussian smoothing. We provide simple analytical results for the three different Wiener filtering techniques 
that have been considered previously in the literature. We also incorporate the optimum estimator 
and its covariance for construction of the corresponding likelihood. The MFs do not fare particularly well in comparison with the optimal
estimators, which are predicted to give much tighter constraints: ${\rm B}_0<0.071$ at $95\%$ CL and  ${\rm B}_0<0.15$ at $99\%$ CL.  These are very close to the predictions 
from the full bispectrum \citep{Hu13}, showing their optimal nature.
We have not considered
the possibility of combining results from different channels which can further improve the constraints. 
\begin{figure}
\begin{center}
{\epsfxsize=10 cm \epsfysize=7 cm {\epsfbox[31 319 586 716]
{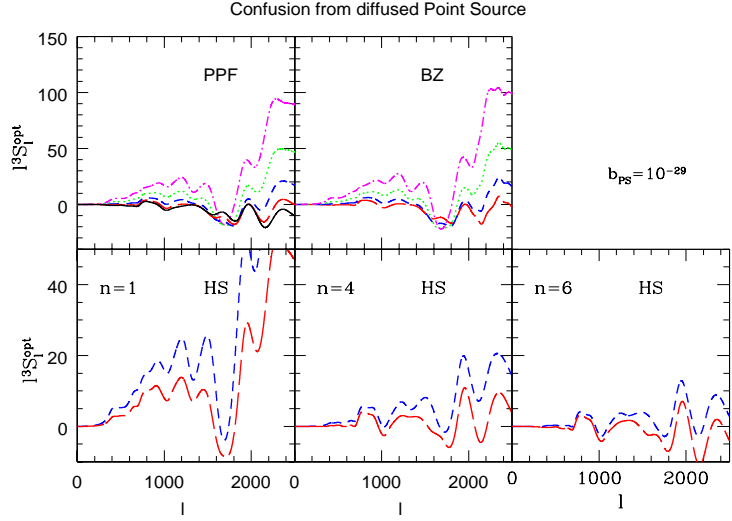}}}
\end{center}
\caption{The confusion from unresolved point sources is plotted for determination of optimum ISW-Lensing skew-spectrum.
The normalisation for point source is fixed at ${\rm b}_{\rm PS}=10^{-29}$. The line-styles used for various models are same as 
that of Figure \ref{fig:clTT}.}
\label{fig:confusion}
\end{figure}

{\bf Contamination from Point Sources and Galactic Foregrounds:} Galactic contamination are a major source of concern
which can affect any study involving the CMB.  They are usually dealt with masking or by using
component separation techniques \citep{Leach08}. The residual bias in the estimation
of primordial non-Gaussianity was found to be small \citep{Hikage08,Komatsu11}.
However, other studies were more conservative in interpreting the results \citep{Chiang03}.
Techniques also exists that involve marginalising over foregrounds \citep{Komatsu02,Komatsu11}.  
Point sources are an additional source of contamination for any study involving MFs. The resolved point sources
with sufficient signal-to-noise can be removed by application of an appropriate mask, but there will be low-flux, unresolved and unsubtracted
sources, comprising radio-galaxies and active galactic nuclei
that emit in radio frequencies through the synchrotron process, and dusty starburst galaxies 
which emit thermally. 
However integrated emission from the Cosmic Infrared Background (CIB) has recently 
been detected by Planck collaboration using the skew-spectrum \citep{Ade:2013dsi}.
Any contamination from unresolved point sources can be estimated using Eq.(\ref{eq:opt}).
Some of the issues involving mask and inhomogeneous noise can be dealt with by computing the {\em cumulant correlators}
that represent MFs in the real-space or in the {\em needlet} basis \citep{MuSmCooReHeCo13}.
The contamination from unresolved point sources (PS) can be estimated using Eq.(\ref{eq:opt}) with $\rm X=ISW$-$lensing$
and $\rm Y=Point\;\; Sources$. The contamination is shown in Figure \ref{fig:confusion}.
For normalisation $b_{\rm PS}=10^{-29}$ the contamination is several orders of magnitude lower
compared to the optimum skew-spectrum depicted in Figure \ref{fig:opt_hs}. We have ignored the
contamination from primordial non-Gaussianity which is expected to be negligible.

\cb{{\bf RS-Lensing and tSZ-Lensing skewspectrum:} The ISW-Lensing cross-correlation at the level of bispectrum has been the focus of our study in this article.
The same techniques can in principle be used to analyse skew-spectra associated 
with the Rees-Sciama(RS)-lensing or thermal Sunyaev Zeldovich(tSZ)-lensing bispectrum to constrain $\rm B_0$.
However, the tSZ-lensing bispectrum depends on detailed modeling of underlying ``gastrophysics'' 
and the S/N of RS-lensing skew-spectrum is below the detection threshold
for ongoing surveys such as the Planck.}

{\bf Beyond the bispectrum:} The results that we have derived here are based on MFs and the optimum
skew-spectrum. Going beyond third-order correlation functions, it is possible to incorporate
the power-spectrum of the lensing potential $\myC^{\phi\phi}_{\ell}$ in constraining $\rm B_0$. Optimized kurt-spectrum
introduced in \cite{Mu11} and later used to analyse 7-year data released by WMAP team \citep{Sm11} can be valuable 
for studies in this direction. These results when combined with results 
from power-spectrum data alone can improve the constraints by an order of magnitude.
The possibility of using polarised CMB maps will be explored elsewhere. 

\cb{{\bf Constraints on $\rm B_0$ from other cosmological data-sets:}Constraints from CMB can provide independent confirmations of constraints derived from 
studies of BAOs, studies of galaxy clusters or that from weak lensing studies, though constraints from galaxy power-spectrum can 
be significantly tighter compared to the constraints derived here $\rm log_{10} B_0 < -4.07$. The scales and redshift probed by
galaxy surveys and CMB observations are very different and are affected by different set of observational systematics.
Hence, these observations play complimentary roles in constraining $\rm B_0$.}

{\bf \cb{Wiener and Wiener-like Filtering and Minkowski Functionals:}} {\em Wiener} and {\em Wiener-like} filtering are generally used for analysing realistic data
to confront issues related to component separation, point-source and galactic masks \citep{Du13}. 
The expressions for MFs in Eq.(\ref{sl1})-Eq.(\ref{sl3}) can be modified by replacing 
the bispectrum by ${\tilde B}_{\ell_1\ell_2\ell_3} = B_{\ell_1\ell_2\ell_3}W_{\ell_1}W_{\ell_2}W_{\ell_3}$.
Various forms of the filters $W_{\ell}$ that were found useful in analysing realistic data are:
$W^{(\rm M)}_\ell= {\myC_{\ell}b^2_{\ell}/\myC^{\rm tot}_{\ell}};$
$W^{(\rm D1)}_\ell= \sqrt{\ell(\ell+1)}{\myC_{\ell}b^2_{\ell}/\myC^{\rm tot}_{\ell}};$
$W^{(\rm D2)}_\ell= {\ell(\ell+1)}{\myC_{\ell}b^2_{\ell}/\myC^{\rm tot}_{\ell}}$.
They correspond to Wiener-filtering ({\rm M}) and Wiener-like filtering using first ({\rm D1}) and second derivatives ({\rm D2}) of the map.
The expression for the covariance can be derived by replacing the power-spectrum by the filtered 
power spectrum $\tilde\myC_{\ell} =W_\ell^2\myC_{\ell}$ in Eq.(\ref{eq:covSij}). By definition, the optimum estimator
includes inverse covariance weighting and its performance cannot be improved by filtering - inclusion of 
weights in the definition of optimum estimator in the numerator and denominator cancel out.
As a final remark, the information content of the skew-spectrum is independent of the power spectrum, as at the lowest order the resulting
cross-correlation will involve five-point spectra which vanish for a Gaussian CMB map.

\section{Acknowledgements}
DM acknowledges support through a STFC rolling grant. DM would like to thank Alexei A. Starobinsky for helpful
discussions. BH and AR are indebted to Sabino Matarrese for useful discussion. 
BH is supported by the Dutch Foundation for Fundamental Research on Matter (FOM).
AR is supported by the European Research Council under the European Community Seventh Framework Programme 
(FP7/2007-2013) ERC grant agreement no. 277742 Pascal.
\label{acknow}
\bibliography{paper.bbl}
\end{document}